\documentstyle[11pt]{article}
\begin{document}
\begin{center}
{\Large\bf Linear Wave Equations and Effective Lagrangians for
Wigner Supermultiplets}
\end{center}
\begin{center}
R. Dahm$^1$ and M. Kirchbach$^2$
\end{center}
\begin{center}
$^1$Institut f\"ur Kernphysik,
Universit\"at Mainz, D--55099 Mainz, Germany\\
(dahm@vkpmzp.kph.uni--mainz.de)\\
$^2$Institut f\"ur Kernphysik,
TH Darmstadt, D--64289 Darmstadt, Germany\\
(kirchb@crunch.ikp.physik.th--darmstadt.de)
\end{center}
\date{\today}

\begin{abstract}
 The relevance of the contracted SU(4) group as a symmetry group of the
pion nucleon scattering amplitudes in the large $N_c$ limit of QCD
raises the problem on the construction of effective
Lagrangians for SU(4) supermultiplets.
In the present study we suggest effective Lagrangians
for selfconjugate representations of SU(4)
in exploiting isomorphism between so(6) and ist universal covering su(4).
The model can be viewed as
an extension of the linear $\sigma$
model with SO(6) symmetry in place of SO(4) and generalizes
the concept of the linear wave equations for particles with arbitrary spin.
We show that
the vector representation of SU(4)
reduces on the
SO(4) level to a complexified quaternion.
Its real part gives rise to the standard linear
$\sigma$ model with a hedgehog configuration
for the pion field, whereas
the imaginary part describes vector meson degrees of freedom
via purely transversal $\rho$ mesons
for which
a helical field configuration is predicted.
As a minimal model, baryonic states
are suggested to appear as solitons of that quaternion.

\end{abstract}

\section{Introduction}
\setcounter{equation}{0}
The large--$N_c$ limit of QCD introduced by t'Hooft \cite{t'Hoo 74}
in the middle of the 70ies as an approximation
scheme to the gauge theory of strong interaction underlies
the idea of the extension of the colour group
from SU($N_c=3$) to SU($N_c>3$). As a direct consequence,
the behaviour of the amplitudes of hadronic processes in the expansion
in powers of $1/N_c$ can be investigated.
Later, $n$--point functions of QCD were systematically
studied by Witten \cite{Wit 79} with the result, that the `large--$N_c$'
scaling of a variety of physical quantities, such as hadronic masses, the
strong meson--baryon vertex and the weak hadron--lepton couplings could be
predicted (so called Witten's $N_c$ counting rules).
Within this treatment,
the meson masses were shown to be
independent of the number of colour degrees of freedom, whereas for
the baryon masses a linear dependence on $N_c$
was obtained (see \cite{Be 93} for a
review). Further,
the weak axial coupling constant $g_A$ of the nucleon
was shown to scale as $O(N_c)$, and
the weak decay constant $f_\pi$ of the pion was found to be of the order
$O(\sqrt{N_c})$. As a consequence, the pseudovector (PV)
$\pi N$ coupling $g_{\pi NN}^{\rm PV} \sim g_A / f_\pi$
was predicted to scale as $O(\sqrt{N_c}$).

With these counting rules the pion--baryon scattering
amplitude was evaluated on the quark level
as independent of $N_c$ in accordance with the unitarity condition
on the S--matrix.
On the level of composite particles,
unitarity is ensured only
if the $\pi N$ scattering amplitude is associated with
direct and crossed Born diagrams including
besides the one--nucleon
also an {\it equal mass} $P_{33}$ intermediate state \cite{Ger 84}
and is mathematically expressed
through vanishing commutator matrix elements  of the type
\begin{equation}
\label{eq:vancomm}
\langle N\mid \lbrack \sigma_a\otimes \tau_b,\,\,
\sigma_k\otimes \tau_l\rbrack \mid
N\rangle
\stackrel{N_c\to \infty}{\longrightarrow}\, 0\,
\end{equation}
in the large $N_c$ limit.
The last equation describes
a Wigner--In\"{o}n\"{u} contraction \cite{Inonu} of the spin--flavour
static group SU(4)
with respect to the nine generators
$\sigma_a\otimes\tau_b$ with $a,b=1,2,3$.
Thus the mass degenerate nucleon and $P_{33}$ states
will constitute the symmetric nonstrange
$\{20\}$--plet which is common
both to the static and the contracted SU(4)
(subsequently denoted by SU'(4)) groups
\cite{Ger 84}, \cite{Da 93}.
Eq.~(\ref{eq:vancomm})
in practice means that in the large
$N_c$ limit $g_A$ and the axial coupling constant
$g_A^{N\to \Delta }$ of the weak $N\to\Delta$--transition have to satisfy the
constraint
\begin{equation}
g_A^2 - {2\over 9} (g_A^{N\to \Delta})^2\,=\,0\,,
\end{equation}
an observation already reported for the case of
some effective models of the nucleon in \cite{Ki 91}.

The relevance of the SU'(4) symmetry for the low energy regime of QCD raises
the question on the extension of the effective models of the
nucleon to such describing Wigner supermultiplets.
In the present study we propose an
effective Lagrangian for selfconjugate meson representations of SU(4). The
model can be viewed as an extension of the linear $\sigma$ model \cite{Go 61}
with SO(6) symmetry in place of SO(4) and generalizes the concept of the
linear wave equations for particles with arbitrary spin \cite{Velo 78}.

The presentation is organized as follows. Sec.~2 reviews the idea for
constructing linear wave equations (LWE) on the foundation of the special
orthogonal group in $5$--dimensional space, SO(5). In sec.~3 we focus on the
isomorphism between the Lie algebras of SO(6)
and its universal covering group
SU(4) and discuss realization in terms of the elements of the
Dirac--Clifford algebra. We suggest SO(6)
 invariant LWE for the selfconjugate
SU(4) irreducible vector representation. Sec.~4 starts with a brief
reminiscence of the linear SO(4)
symmetric $\sigma$ model \cite{Go 61}. After
that an SO(6) invariant effective meson Lagrangian is suggested.
In sec. 5 discussion on the perspectives for SU(4) fermionic
states description is given.
The paper ends with a short summary.

\section{Linear wave equations}
\setcounter{equation}{0}
The general ansatz for linear relativistic wave equations has the form
\cite{Velo 78}
\begin{equation}
\label{eq:gleins}
(\alpha_\mu\partial^\mu+\chi\, 1_{\rm n\times n})\Psi_{\{r\}}(x)\,=\,0
\end{equation}
with indices $\{r\}=\{1,...,n\}$, $\mu\,=\,0,1,2,3$.
The field $\Psi_{\{r\}}(x)$ denotes
a multicomponent vector state transforming as an
$n$--dimensional irreducible representation
(irrep) of the Lorentz group and $\chi$ is a constant related to the mass.
The requirement on Lorentz invariance of eq.~(\ref{eq:gleins}) leads to the
commutation relation \cite{Ro 60}
\begin{equation}
\label{eq:glzwo}
[S_{\mu\nu },\alpha_\eta]=\alpha_\mu g_{\nu\eta}-\alpha_\nu g_{\mu\eta}
\end{equation}
between the ${n\times n}$--matrices $\alpha_\mu$ and the six generators
$S_{\mu \nu }$ of the homogeneous Lorentz group. The quantities $S_{\mu\nu}$
satisfy the Lie algebra
\begin{eqnarray}
\label{eq:glzwoa}
&&[S_{\mu\nu},S_{\rho\sigma}]\nonumber\\
& = &- g_{\mu\rho}S_{\nu\sigma}+ g_{\nu\rho}S_{\mu\sigma}
+ g_{\mu\sigma}S_{\nu \rho}- g_{\nu\sigma}S_{\mu\rho}
\end{eqnarray}
with $g_{\mu\nu}\,=\,(1,-1,-1,-1)$.
The algebra of the homogeneous
Lorentz group in (3+1) space--time dimensions was
extended to that in (4+1) space--time dimensions
by Pauli \cite{Pa 33} for the purpose of Kaluza--Klein theory.
Pauli was the first who emphasized already in 1933 the relevance of
SO(5) for relativistic problems.
The Lie algebra of the SO(5) group\footnote{We follow the discussion
(\protect\cite{Bh 45}, \protect\cite{Fi 74}, \protect\cite{MMN}) given in
terms of compact groups and comment later on appropriate noncompact
groups.} is obtained by
completing eq.~(\ref{eq:glzwoa}) by the following commutation relations:
\begin{equation}
\label{eq:glvier}
[l_{\mu},l_{\nu}]=S_{\mu\nu}\,,\quad
[S_{\mu\nu},l_{\sigma}]=g_{\nu\sigma}l_{\mu}-
g_{\sigma\mu}l_{\nu}\,.
\end{equation}

Later, in 1945, Bhabha \cite{Bh 45} observed that
Eq.~(\ref{eq:glzwo}) is satisfied if the matrices $\alpha_\mu$ are identified
with the four SO(5) generators $l_{\mu}$ as
\begin{equation}
\alpha_{\mu}=l_{\mu}\,.
\end{equation}
Insertion of eq.~(\ref{eq:glvier}) into eq.~(\ref{eq:glzwo}) leads to the
algebra
\begin{equation}
[[\alpha_{\mu},\alpha_{\nu}],\alpha_{\eta}]=
g_{\nu\eta}\alpha_{\mu}-g_{\mu\eta}\alpha_{\nu}\,.
\end{equation}
The vector fields $\Psi_{\{r\}}(x)$ in eq.~(\ref{eq:gleins}) therefore
can be viewed as irreps of SO(5),
the special orthogonal group in five dimensions
\cite{Fi 74}.
First order wave equations for
scalar and vector fields are associated with
the 5-- and 10--dimensional
SO(5) irreps and are usually referred to as Duffin--Kemmer--Petiau equations
\cite{Ke 39}.
Particles of arbitrary spin are related to higher dimensional
SO(5) multiplets and the corresponding equations are often
called `Bhabha equations'.
For more details the interested reader is referred
to the most extensive study on this subject performed in the series of papers
by Krajcik and Nieto \cite{MMN}.

To construct the Lagrangian underlying eq.~(\ref{eq:gleins}) it is necessary
to define a conjugation operation on the fields $\Psi_{\{r\}}(x)$. This is
done by means of a $n\times n$ matrix $\eta$ with the property
\begin{equation}
-D\left(\Lambda\right) \eta = \eta D\left(\Lambda\right)^T\, .
\end{equation}
Here, $D\left(\Lambda\right)$ stands for the $n\times n$ matrix representations
of the Lorentz transformations. In the special case of the four dimensional
irrep of SO(5) the Bhabha equation
is identical to the Dirac equation and
the matrix $\eta$ can be expressed by $\alpha_0$ as $\eta =2\alpha_0$
\cite{MMN}. It is easily proved that the quantity $\overline{\Psi}(x)\Psi(x)$=
$\Psi^{+}(x)\eta\Psi(x)$ transforms as a Lorentz scalar. Thus the
Lagrangian leading to eq.~(\ref{eq:gleins}) can be written as
\begin{equation}
{\cal L} =\overline{\Psi}(x)\partial\cdot\alpha\Psi(x)
+\chi\overline{\Psi}(x)\Psi(x)\,.
\end{equation}
The irreducible representations of the group SO(5) as introduced above
can be used for the description
of one--flavour states only and thus Bhabha's
equations are not directly
applicable as wave equations for the spin--flavour multiplets
emerging in the large $N_c$ limit of QCD. Nevertheless, the trail blazed by
Bhabha's equations can be pursued and generalized to incorporate
isospin degrees of freedom. This will be the subject of the next section.
There we will show that SO(6) invariant LWE can be used to describe the
selfconjugate SU(4) spin--flavour $\{15\}$--plet
by means of the chain SU(4) $\sim$ SO(6) $\supset$ SO(5).

\section{SO(6) invariant linear wave equations and isospin degrees of
freedom}
\setcounter{equation}{0}
To incorporate the isospin deg\-rees of freedom into the LWE we
exploit the
isomorphism (denoted by $\simeq$) between the Lie algebras su(4) and so(6).
As is well known from group theory \cite{Cornwell}, \cite{Gil}, a doubly
connected orthogonal group SO($n$) has a simply connected universal covering
group denoted by $Spin(n)$. The four cases in which the $Spin$--groups
correspond to classical groups are \cite{Cornwell}, \cite{Gil}
\begin{eqnarray}
\label{eq:drei}
Spin(3) & \simeq & {\rm SU}(2)\,,\\
\label{eq:neunull}
Spin(4) & \simeq & {\rm SU}(2)\,\otimes\,{\rm SU}(2)\,,\\
\label{eq:eins}
Spin(5) & \simeq & {\rm Sp}(2)\,,\\
\label{eq:zwei}
Spin(6) & \simeq & {\rm SU}(4)\,,
\end{eqnarray}
whereas for the Lie algebras the isomorphisms
\begin{eqnarray}
\label{eq:neueins}
{\rm su(2)} & \simeq & {\rm so}(3)\,,\\
\label{eq:neuzwei}
{\rm su}(2)\,\oplus\,{\rm su}(2) & \simeq & {\rm so}(4)\,,\\
\label{eq:neudrei}
{\rm sp}(2) & \simeq & {\rm so}(5)\,,\\
\label{eq:neuvier}
{\rm su}(4) & \simeq & {\rm so}(6)
\end{eqnarray}
hold.

The first three equations have relevant physical applications. For example,
eq.~(\ref{eq:neueins}) leads to equivalent
description of rigid body rotation
in three dimensional space in terms of the Euler--angles and
the Cayley--Klein parameters, respectively.
Eq.~(\ref{eq:neuzwei}) underlies the
construction of effective models in Chiral Dynamics like the $\sigma$ model
\cite{Go 61}, whereas eq.~(\ref{eq:neudrei}) builds the basis for
the linear wave equations \cite{Velo 78}.
The aim of the present study is to associate a physical model with
eq.~(\ref{eq:neuvier}).

Among eqs.~(\ref{eq:drei})--(\ref{eq:zwei}) the first and fourth
are the most fundamental ones
because the corresponding algebras are in addition isomorphic to the ones
generated by the elements of the Clifford algebras $C_2$ and $C_4$,
respectively. Indeed, it was shown by Barut \cite{Ba 64} that the Lie group
generated by the 15 elements of the Dirac--Clifford algebra $C_4$ is
isomorphic to the six dimensional real Lorentz group with the metric
(-1,-1,+1,-1,-1,-1),
and is thus related to the compact group SO(6)\footnote{The Clifford
algebra has actually 16 elements but one of them commutes with all the others
and equals the identity operator.}. To see this, one has first to consider the
antisymmetric set of generators
\begin{displaymath}
{\bf S_{ab}}=
\left(
\begin{array}{cccccc}
0&\gamma_5&-\gamma_5\gamma_0&\gamma_5\gamma_1&\gamma_5\gamma_2&
\gamma_5\gamma_3\\
&0 &-\gamma_0&\gamma_1&\gamma_2&\gamma_3\\
& & 0& \gamma_0\gamma_1&\gamma_0\gamma_3&\gamma_0\gamma_3\\
& & & 0& \gamma_2\gamma_1 &\gamma_3\gamma_1\\
& & & & 0& \gamma_3\gamma_2\\
& & & & & 0 \\
\end{array}\right)
\end{displaymath}
with $a,b=1,...,6$ and then to re-express them by the physically
more convenient generators,
\begin{eqnarray}
S_{\mu\nu } & = & {1\over 2}(\gamma_\mu \gamma_\nu -g_{\mu\nu})\,,\\
l_\mu & = & {1\over 2}\gamma_\mu\,,\\
\tilde{l}_\mu & = & {1\over 2}\gamma_5\gamma_\mu\,,\\
K & = & {\gamma_5\over 2}\,,\qquad \mu = 0,1,2,3 \,.
\end{eqnarray}

It is easy to prove the commutation relations
\begin{eqnarray}
\lbrack S_{\mu\nu},\tilde{l}_{\sigma}\rbrack & = &
g_{\nu\sigma } \tilde{l}_\mu -g_{\sigma \mu }\tilde{l}_\nu \,,\\
\lbrack\tilde{l}_\mu,\tilde{l}_\nu\rbrack & = & S_{\mu\nu }\,,\\
\lbrack l_\mu ,\tilde{l}_\nu \rbrack & = & -g_{\mu\nu} K\,,\\
\lbrack K,l_\mu \rbrack & = & \tilde{l}_\mu \,,\\
\lbrack K,l_\mu \rbrack & = & -l_\mu\,,\\
\lbrack K, S_{\mu\nu}\rbrack & = & 0\,.
\end{eqnarray}
In joining them to eqs.~(\ref{eq:glzwoa})--(\ref{eq:glvier}),
the Lie algebra of the group SO(6) is obtained.
For this reason, the matrices $\alpha_\mu$
entering eq.~(\ref{eq:gleins}) can be viewed as the generators $l_\mu$ of
the group SO(6), and the field representations
$\Psi_{\{r\}}(x)$ will behave
as irreps belonging to SO(6). It was Barut \cite{Ba 64} who showed that the
SO(6) generators $S_{12}$, $-iS_{13}$, and $-iS_{23}$
span a new rotation group disjoint from the spin group. This former
group can be identified with the group of isospin.
Thus the advantage of SO(6)
symmetric LWE will be that their solutions can have spin--flavour content.
Since only selfconjugate irreps of su(4)
can be related directly to so(6) irreps, the SO(6)
invariant LWE can be employed only for mesonic spin--flavour supermultiplets
such as the selfconjugate $\{15\}$--plet. The linear relativistic SO(6)
equation for the (SO(5) reducible) $\{15\}$--plet reads:
\begin{equation}
(\partial_\mu\alpha^\mu_{[15\times 15]} +
\chi\,1_{[15\times 15] })
\Psi_{\{15\}}(x)\,=\,0\,.
\end{equation}
The decomposition of the $\{15\}$--plet
within the SO(5) basis to which the Lorentz transformation
directly applies, reads:
\begin{equation}
\{15\}=\{5\}\oplus\{10\}\,.
\end{equation}
Note, that the DKP algebra ${\cal B}(1)$ \cite{Fi 74} is reducible into the
two inequivalent representations of dimensions 5 and 10, respectively,
which are used for the description of
spin--0 and spin--1 particles \cite{Ke 39},\cite{MMN}.
For on shell particles the LWE for the $\{5\}$-- and
$\{10\}$--plets are equivalent to the Klein--Gordon  and the
Proca equations, respectively.
Thus the so(5) decomposition of
the so(6) vector $\{15\}=\{5\}\oplus\{10\}$ fits naturally
into the DKP algebra.

\noindent
The SU(4) particle content associated with that states will be\footnote{
The relevant degrees of freedom for so(6) are actually
the real cartesian components.
It is the linear character of the representations in
combination with the linear wave equations which
makes it possible to use
equal dimensional
multiplets containing complexified (i.e. charged) fields.}
\begin{eqnarray}
\label{eq:arreins}
\{5\} & = & col(\omega_3 ,\rho^0_3, \pi^{+},\pi^{-},\pi^{0})\\
& \stackrel{{\rm so}(4)}{\longrightarrow} &
col(\rho^0_3, \pi^{+},\pi^{-},\pi^{0})\oplus\omega_3,\nonumber
\end{eqnarray}
\begin{eqnarray}
\label{eq:arrzwei}
\{10\} & = & col(\rho^{+}\!\uparrow,\rho^{-}\!\uparrow,
\rho^{0}\!\uparrow,
\rho^{+}\!\downarrow,\rho^{-}\!\downarrow,
\rho^{0}\!\downarrow,
\rho^{+}_3,\rho^{-}_3,\omega\!\uparrow,\omega\!\downarrow)\\
& \stackrel{{\rm so}(4)}{\longrightarrow} &
col(\rho^{+}\!\uparrow,\rho^{-}\!\uparrow,
\rho^{0}\!\uparrow,\rho^{+}\!\downarrow,
\rho^{-}\!\downarrow,\rho^{0}\downarrow)
\,\oplus\,col (\rho^{+}_3,\rho^{-}_3,
\omega\!\uparrow,\omega\!\downarrow)\, .
\end{eqnarray}
Now the idea is to interprete the spin--flavour $\{5\}$--plet
in (\ref{eq:arreins}) as the solution of the standard DKP-equation for
a scalar particle field
$\psi_{DKP}(p)=(\chi /p_0V)^{1/2}{\cal U}_{DKP}^{(5)}(p)e^{ip\cdot x}$
 \cite{Ro 60}, \cite{MMN}
which leads to the correspondence
\begin{equation}
\label{eq:brauchfuenf}
{\cal U}^{(5)}_{DKP}(p)=(2\chi^2 )^{-1/2}
\left(
\begin{array}{c}
-\chi \phi\\
\partial_0\phi\\
\partial_1\phi\\
\partial_2\phi \\
\partial_3\phi
\end{array}\right) =
(2\chi^2)^{-1/2}
\left(
\begin{array}{c}
\omega_3\\
\rho^0_3\\
\pi_x\\
\pi_y\\
\pi_z\\
\end{array}
\right)\, .
\end{equation}
Here the Lorentz indices run from $\mu =0,...,3$, whereas
the coordinates in intrinsic isospin space are denoted
by $x,y,z$.
Similarly, the solution of the Duffin--Kemmer--Petiau
equation for a vector particle field (denoted by $a_\mu$)
\begin{equation}
\label{eq:zehner}
{\cal U}^{(10)}_{DKP}(p)\, =
(2\chi^2)^{-1/2}
\left(
\begin{array}{c}
-\partial_1 a_0 - \partial_0 a_1\\
-\partial_2 a_0 - \partial_0 a_2\\
-\partial_3 a_0 - \partial_0 a_3\\
\partial_2 a_3-\partial_3 a_2\\
\partial_3 a_1-\partial_1 a_3\\
\partial_1 a_2 - \partial_2 a_1\\
-\chi\, a_1\\
-\chi\, a_2\\
-\chi\, a_3\\
-\chi\, a_0
\end{array}
\right)\equiv
(2\chi^2)^{-1/2}
\left(
\begin{array}{c}
e_1\\
e_2\\
e_3\\
h_1\\
h_2\\
h_3\\
-\chi a_1\\
-\chi a_2\\
-\chi a_3\\
-\chi a_0
\end{array}
\right )
\end{equation}

\noindent
can be used to predict a correlation between
spin--flavour and space--time degrees of freedom for the vector mesons.
The isotriplet helicity doublet
$\lbrace \rho^{+,-,0}\uparrow \rbrace\oplus
\lbrace\rho^{+,-,0}\downarrow\rbrace $
ist most naturally mapped onto
the  $\lbrace 1,0\rbrace\oplus\lbrace 0,1\rbrace $
representation of so(4),
whereas the charge/spin doublets $\lbrace \rho^+_3,\rho^-_3\rbrace
\oplus \lbrace \omega\uparrow,\omega\downarrow\rbrace $
are mapped onto $\lbrace1/2,1/2\rbrace$ according to
\begin{equation}
\label{eq:brauchzehn}
\left(
\begin{array}{c}
\rho^+\uparrow\\
\rho^-\uparrow\\
\rho^0\uparrow\\
\rho^+\downarrow\\
\rho^-\downarrow\\
\rho^-\downarrow\\
\end{array}
\right) =
\left(
\begin{array}{c}
e_1 + ih_1\\
e_2 + ih_2\\
e_3 +ih_3\\
e_1-ih_1\\
e_2-ih_2\\
e_3-ih_3\\
\end{array}
\right)\, ,
\end{equation}
and
\begin{equation}
\left(
\begin{array}{cc}
\omega\uparrow & \rho^-_3\\
\rho^+_3 & \omega\downarrow
\end{array}
\right)\to -\chi
\left(
\begin{array}{cc}
 a_0 +a_3 & a_1 -ia_2\\
a_1 +i a_2 & a_0-a_3
\end{array}
\right)\,.
\end{equation}

In eq.~(\ref{eq:zehner}), ${\cal U}^{(10)}_{DKP}(p)$ stands for a massive
$DKP$ spinor field related to the solution $\Psi_{DKP}$
of the spin--1 DKP equation via
$\Psi_{DKP}=\exp(\chi /p_0V)^{1/2}{\cal U}_{DKP}^{(10)}(p)$
\cite{Ro 60}, \cite{MMN}.
The mass spectrum of the irreps of the group SO(5) was studied
in great detail in \cite{MMN}. There, it was shown that after
SO(4)--reduction of a quintuplet a quadruplet
($col(\rho^0_3,\pi^+,\pi^-,\pi^0)$ in our case) of finite
mass, $\chi$, and a singlet
($\psi_{\lbrace 0\rbrace}=\omega_3$ in our case)
satisfying at rest the equation
$(0\cdot \partial_t^2 -\chi^2)\psi_{\lbrace 0\rbrace }=0$
and therefore of infinite mass, appear.
Similar analysis shows that the 6--dimensional vector
$col(\rho^+\uparrow ,\rho^-\uparrow ,\rho^0\uparrow ,
\rho^+\downarrow ,\rho^-\downarrow ,\rho^0\downarrow )$
will be of finite mass $\chi$, whereas
an infinite mass should be attributed to the accompanying four--vector
($col (\rho^+_3,\rho^-_3,\omega\uparrow ,\omega\downarrow$ ) in our case).
The latter state,
the only one for which spin and isospin degrees of freedom do
not decouple, is expunged from the low mass region
and becomes a non--observable degree of freedom
for the low energy hadron physics considered here.
Thus the $\omega $ meson as well as the longitudinal modes
of the positively and negatively charged $\rho$ mesons
drop out trough SO(6) $\supset$ SO(5) $\supset$ SO(4) reduction of
the defining representation.
Note, that in the case of pure flavour SU(4) this
reduction scheme would be less natural, for the
decomposition of the purely
pseudoscalar flavour $\lbrace 15\rbrace$--plet
into SO(4) irreps requires additional information.

In the next section we present ideas on how to construct effective
mesonic field theories by means of
the SO(6) invariant equations developed above.

\section{Effective SO(6)--invariant Lagrangian for the
spin-fla\-vour $\{15\}$--plet}
\setcounter{equation}{0}
The most famous effective theory in low energy hadron physics is the linear
$\sigma$ model. It is based on a pure mesonic Lagrangian of the type $\phi^4$
\cite{Go 61} which is the maximal renormalizable theory in (3+1) space--time
dimensions. The underlying symmetry of the model is the chiral group
SU(2) $\otimes$ SU(2) acting as the universal covering group of SO(4) (compare
eq.~(\ref{eq:neunull})). For this purpose a relativistic, SO(4) invariant
Lagrangian is constructed in terms of the four vector
$\lbrace 1/2,1/2\rbrace\,\equiv\,\psi_{\lbrace 4\rbrace}$ in place
of the scalar $\phi$ as follows:
\begin{eqnarray*}
{\cal L}_\sigma (x) & = & {1\over 2}
\partial^{\mu}\, \psi_{\lbrace 4\rbrace }(x)\partial_{\mu}\,
\psi_{\lbrace 4\rbrace }(x)- {\cal V}(x)\,,\\
{\cal V}(x) & = & {1\over 2}\,\mu^{2}||\psi_{\lbrace 4 \rbrace }(x)||^{2}
-{1\over 4}\,\lambda^2 ||\psi_{\lbrace 4\rbrace}(x)||^{4}
\end{eqnarray*}
with the constraint
\begin{equation}
\label{eq:chircircconstr}
||\psi_{\lbrace 4\rbrace }||^{2}\,=\,\sigma^{2}(x)+\vec{\pi}^{2}(x)\,=f^2\,.
\end{equation}
\noindent
This Lagrangian is related to unitary theories by use of a nonlinear
realization
\begin{equation}
||\psi_{\{4\}}||^{2}\,=\,{f^2\over 2}\,tr\left(U^{+}(x)U(x)\right)\,,
\end{equation}
\begin{equation}
U(x)\,=\,\exp\left(-i\vec{\tau}\cdot{\vec{\varphi}(x)}\right)
\,=\,{1\over f}\left(\sigma(x)1-\sum_{j=1}^{3}\pi_{j}(x)\,i\tau_{j}\right)\,.
\end{equation}
Here, $\vec{\varphi}(x)$ stands for a dimensionless parameter set
termed to as `generalized chiral angles' which allows for the
parametrization
\begin{equation}
\label{eq:arrdrei}
{\sigma\over f}=\,\cos\varphi\,,\quad
{\pi_{j}\over f} =\,\frac{\varphi_{j}}{\varphi}\sin\varphi\,,
\quad\varphi\,=\,|\vec{\varphi}|
\end{equation}
whereas $f$ rescales the norm of $\psi_{\lbrace 4\rbrace}$ to unity.
The interpretation of this nonlinear realization and its transformations
in Chiral Dynamics becomes much more apparent when the field $U(x)$ is
interpreted in terms of real quaternions\footnote{Subsequently, we denote
real, complex, quaternionic and octonionic numbers by {\bf R}, {\bf C},
{\bf H} and {\bf O}, respectively.}. Embedding quaternions by Pauli
matrices $\tau_{j}$ into two dimensional complex spaces,
\begin{equation}
q_{0}\,=\,1_{[2\times 2]}\,,q_{j}\,=\,-i\,\tau_{j}\,\Longleftrightarrow\,
U(x)\,=\,{1\over f}
\left(\sigma(x) q_{0} + \sum_{j=1}^{3}\,\pi_{j}(x)\,q_{j}\right)\,,
\end{equation}
$U(x)\in$ USp(2) $\simeq$ U(1,$q$) denotes a normalized real quaternion.
Transformations of $U(x)$ may be performed just by quaternionic
multiplication, i.e. for $a,b\in$ U(1,$q$) we have
\begin{equation}
\label{eq:chirtrafo}
U\longrightarrow U'\,=\,a\,U\,b^{+}\,,
\end{equation}
where `$^{+}$' denotes quaternionic conjugation $q_{0}\to q_{0}$,
$q_{j}\to -q_{j}$.
The transformation (\ref{eq:chirtrafo}) may be parametrized by six real
parameters $\vec{\epsilon}_{R}$ and $\vec{\epsilon}_{L}$ according to
\begin{equation}
a\,=\,\exp\left(-i\vec{\tau}\cdot{\vec{\epsilon}_{R}}\right)\,,\qquad
b^{+}\,=\,\exp\left(i\vec{\tau}\cdot{\vec{\epsilon}_{L}}\right)\,,
\end{equation}
so that the relation to right/left transformations of the meson field
$U$ used in Chiral Dynamics is
obvious. The conservation of the norm of $U$ can either be calculated
directly by use of eq.~(\ref{eq:chirtrafo}) or from the property of the
real quaternions to form a division algebra, i.e.
\begin{equation}
||U'||\,=\,||a\,U\,b^{+}||\,=\,||a||\,||U||\,||b^{+}||\,=\,||U||\,.
\end{equation}
Furthermore, the restriction $\epsilon_{R}\,=\,\epsilon_{L}$ leads to $a=b$
and the transformation
\begin{equation}
U\longrightarrow U'\,=\,a\,U\,a^{+}
\end{equation}
of the field $U$. Thus, SU(2) flavour as the `diagonal subgroup' of
Chiral Dynamics emerges automatically as automorphism group of real
quaternions \cite{finkel}. The Lagrangian of the linear sigma model
as expressed in terms of the U--field reads
\begin{equation}
{\cal L}_\sigma =
-{f^2\over 4} tr \left(\partial_\nu U(x)\partial^\nu U(x)\right)
-{1\over 2}\mu^2 ||U(x)||^{2} +{1\over 4}\lambda^{2}||U(x)||^{4}\,.
\end{equation}
It is noteworthy, that the properties discussed above are special
features of the $\lbrace 1/2,1/2\rbrace$ irrep which are in general
not shared by other representations of the orthogonal group.

\noindent
In the Skyrme model \cite{Sky 62} where the Lagrangian
of the linear sigma model is completed by
the so called Skyrme term \cite{Sky 62}
the nucleon is described
as a soliton of the 2$\times$2 quaternion field U(x)
realizing at fixed times a map of the
unit three space sphere $S^3$ from {\bf R}$^4$ onto the SU(2)
group manifold $S^3$ (see \cite{Zah 86} for a review).
Once ${\cal L}_\sigma$ has been written in terms of
an exponentially represented SU(2) group element, its
generalization to higher SU(N$_F\, > 2 $) can be considered
with the drawback that the group manifolds may not longer be spheres.
Such (complex) generalizations of the Skyrme model have extensively been
considered in the literature (see \cite{Wal 93} for a recent work).

Generalizing the symmetry of the linear sigma model from SO(4) to SO(6)
with the 15--dimensional defining representation of so(6) in place of
$\lbrace 1/2,1/2\rbrace $, it is necessary to emphasize the important
r\^{o}le of quaternions and their embedding in complex vector spaces.
The embedding of twofold quaternions leads to USp(4) $\simeq$ U(2,$q$)
and the group SU(4). The vector
representation of the algebra su(4) is cast into the
traceless second rank tensor $M^A_B(x)$, $1\leq A,B\leq 4$
according to
\[
M^A_B(x)\,=\left(
\begin{array}{cccc}
{1\over 2}(\pi^{0}+\omega_{3}+\rho^{0}_{3}) &
\sqrt{{1\over 2}}(\pi^{+}+\rho_{3}^{+}) &
\sqrt{{1\over 2}}(\omega\uparrow +\rho^{0}\uparrow) & \rho^{+}\uparrow\\
\sqrt{{1\over 2}}(\pi^{-}+\rho^{-}_{3}) &
{1\over 2}(-\pi^{0}+\omega_{3}-\rho_{3}^{0}) &
\rho^{-}\uparrow & \sqrt{{1\over 2}}(\omega\uparrow -\rho^{0}\uparrow)\\
\sqrt{{1\over 2}}(\omega\downarrow + \rho^{0}\downarrow) &
\rho^{+}\downarrow & {1\over 2}(\pi^{0}-\omega_{3}-\rho^{0}_{3}) &
\sqrt{{1\over 2}}(\pi^{+}-\rho_{3}^{+})\\
\rho^{-}\downarrow & \sqrt{{1\over 2}}(\omega\downarrow -\rho^{0}\downarrow) &
\sqrt{{1\over 2}}(\pi^{-}-\rho^{-}_{3}) &
{1\over 2}(-\pi^{0}-\omega_3 +\rho^{0}_3)
\end{array}
\right)
\]
The exponential representation of the meson supermultiplet is given by
\begin{equation}
\label{eq:dakisoli}
U(x)=\exp\left(iM(x)\right)
\end{equation}
and describes finite transformations with respect to SU(4).
If we take into account the isomorphism su(4) $\longleftrightarrow$ so(6),
it is possible to relate the selfconjugate representations of su(4) to
so(6) and treat them on the same footing. Towards the construction of
a mesonic SO(6) invariant Lagrangian the following ideas can be exploited:

\begin{enumerate}
\item  The first idea relies on the property of the unitary/orthogonal
groups to preserve the norms of the corresponding
irreducible representations.
Because of the isomorphism between the
su(4) and so(6) algebras mentioned above,
the selfconjugate su(4) tensor $M^A_B$
is associated with the vector representation of so(6) and the following
effective Lagrangian can be written:
\begin{equation}
\label{eq:arrvier}
{\cal L}\,=\,\overline{\Psi}_{\{15\}}\alpha\cdot\partial\Psi_{\{15\}}
+\mu^{2}\,\overline{\Psi}_{\{15\}}\eta\Psi_{\{15\}}
-\lambda^{2}\left(\overline{\Psi}_{\{15\}}\eta\Psi_{\{15\}}\right)^{2}\,.
\end{equation}
Note, that we have omitted the space--time argument $x$
to simplify the notation.

The constants $\mu^2$ and $\lambda^{2}$ have to be determined by comparison
to suitably chosen experimental data.
{}From the latter equation it follows that if we neglect the
$\lbrace 10\rbrace$--plet, the interaction term
reduces to that of the standard sigma model with the
$\rho^0_3$ field appearing in place of the $\sigma$ meson.
Now the correspondence
between the spin--flavour and space--time
degrees of freedom given in eq.~(\ref{eq:brauchfuenf})
shows that for the case of a radial configuration
of the scalar field $\phi(r)$
the famous hedgehog ansatz \cite{Zah 86}
of the nonlinear sigma model
$\pi_x(r) \sim \vec{r}_1$, $\pi_y(r)\sim \vec{r}_2$,
and $\pi_z (r)\sim \vec{r}_3$,
is recovered.

In a similar way, eq.~(\ref{eq:brauchzehn}) will lead to a helical
field configuration for the
vector mesons.
Detailed study of the solution of the equations proposed will be
given in a forthcoming paper.

\item The second possibility is to introduce nonlinear terms along the
line of conformally invariant spinor equations (see \cite{Fu 87} for a
review) via
\begin{equation}
\label{eq:arrfuenf}
\left(\alpha\cdot\partial\,
+\,F(\overline{\Psi}_{\{15\}},\Psi_{\{15\}})\right)\Psi_{\{15\}}\,=\,0\,,
\end{equation}
with
\begin{eqnarray}
F(\overline{\Psi}_{\{15\}},\Psi_{\{15\}}) & = & F_{1}\,+\,F_2\,\eta\,+\,
F_{3}\,\alpha^{\mu}
\left(\overline{\Psi}_{\{15\}}\eta\alpha_{\mu}\Psi_{\{15\}}\right)\nonumber\\
& & +\,F_{4}\,S^{\mu\nu}
\left(\overline{\Psi}_{\{15\}}\eta S_{\mu\nu}\Psi_{\{15\}}\right)\,,\\
S_{\mu\nu} & = & {i\over 4}
\left(\alpha_{\mu}\alpha_{\nu}-\alpha_{\nu}\alpha_{\mu}\right)\,.\nonumber
\end{eqnarray}
Here $F_{1}$, $F_{2}$, $F_{3}$ and $F_{4}$ stand for some arbitrary scalar
functions of $\overline{\Psi}\Psi$ and $\overline{\Psi}\eta\Psi$.
The proof that such equations are solvable can be found i.e. in \cite{Fu 87}.

\item A third effective model can be constructed by means of the field
${\cal U}(x)$ determining the norm of M in the reduction scheme SO(6)
$\supset$ SO(5) $\supset$ SO(4):
\begin{equation}
{\cal U}(x)={1\over F}\left(u_{1}(x)\,+\,iu_{2}(x)\right)\,,
\end{equation}
with the quaternions $u_{1}$ and $u_{2}$ given by:
\begin{eqnarray}
u_{1}(x) & = & \rho^0_3(x)\left(
\begin{array}{cc}
1_{2} & 0\\
0 & -1_{2}
\end{array}\right)
-\vec{\pi}(x)\cdot\left(
\begin{array}{cc}
-i\vec{\tau} & 0\\
0 & -i\vec{\tau}
\end{array}\right)\,,\nonumber\\
u_{2}(x) & = &
-\sqrt{2}\left(
\begin{array}{cc}
0 & -i\vec{\tau}\cdot\vec{\rho}(x)\!\uparrow\\
-i\vec{\tau}\cdot\vec{\rho}(x)\!\downarrow & 0\\
\end{array}
\right)\,.
\end{eqnarray}
The new constant $F$ instead of $f$ rescales the norm of the meson
supermultiplet to unity. It is then natural to suppose that the fermionic
SU'(4) $\{20\}$--plet, occurring in the large $N_c$ limit of QCD, could
emerge as a soliton of the ${\cal U}(x)$ field.
\end{enumerate}
Within the schemes proposed above meson--meson scattering inclusive
corrections from $\{15\}$--loops can be calculated.

\noindent
Effective Lagrangians with linear kinetic terms have been
considered e.g. in ref. \cite{Ok 79} where
an SO(5) invariant Duffin--Kemmer-Petiau Lagrangian has
been used for the construction of a nonlinear
sigma model. There, four different SO(5) quintuplets have been
in turn associated with the $\pi^+$, $\pi^-$, $\pi^0$ and
the $\sigma$ meson fields. Furthermore,
an auxiliary SO(5) singlet state has been introduced and exploited
as a Lagrange multiplier
to account for the chiral circle constraint of (\ref{eq:chircircconstr}).
This field has then been organized together with
the four quintuplets mentioned above
into a reducible $\{21\}$--plet undergoing
SO(5) algebra transformations. Note that {\it no}
SU(4) representation could be mapped onto this $\{21\}$--plet.
Our approach differs principally from the one presented
in \cite{Ok 79} since it is based on {\it irreducible}
representations {\it common} both to SO(6) and its covering group
SU(4).

\section{\bf Perspectives for SU(4) fermion states description }
\setcounter{equation}{0}
To include explicit fermionic degrees of freedom
in the Lagrangians, one has to construct spinor representations of SU(4),
the universal covering group of SO(6). To benefit once more
from quaternions one can exploit isomorphism
between SO(5) and USp(4),
the maximal subgroup in SU(4),
which allows one to consider
the stereographic projection
$S^4$ from ${\bf R}^5$ onto ${\bf H}$
in analogy to the stereographic projection
$S^{2}$ $\to$ {\bf C}.

Via the stereographic projection each point of
the unit sphere $S^{2}$ of {\bf R}$^{3}$
placed on a line passing, say, through the north pole,
is mapped onto the complex number $z$
determining the intersection between that line and
the equatorial plane \cite{gelf}.
Introducing complex homogeneous coordinates $z_{1}$ and $z_{2}$
via $z\longrightarrow z_{1}/z_{2}$, it is observed that
M\"obius transformation of the complex plane
\begin{equation}
z'\,=\,\frac{\alpha z+\beta}{{-\beta^{*}} z+\alpha^{*}}
\,=\,\frac{\alpha z_{1}+\beta z_{2}}{{-\beta^{*}}z_{1}+\alpha^{*}z_{2}}
\end{equation}
with $|\alpha|^{2}+|\beta|^{2}=1$
are equivalent to SU(2) transformations of the two--component vectors $\psi$,
\begin{equation}
\psi\,=\,\left(\begin{array}{c}z_{1}\\z_{2}\end{array}\right)\,\longrightarrow\,
\psi'\,=\,\left(\begin{array}{c}z'_{1}\\z'_{2}\end{array}\right)\,=\,
\left(
\begin{array}{cc}
\alpha & \beta\\
-\beta^* &\alpha^*
\end{array}
\right)
\left(\begin{array}{c}z_{1}\\z_{2}\end{array}\right)\, ,
\end{equation}
and thus describe rotations of $S^2$.
In generalizing the M\"obius transformations
in the complex plane to Sl(2,$c$), particles of arbitrary spin can be
described by means of higher rank tensors (so called spin--tensors)
constructed as direct products of the spinors $\psi$ and $\psi^{+}$
with the property to satisfy the Bargmann--Wigner equations.

\noindent
In a similar way, the points of the unit sphere $S^4$ of {\bf R}$^5$
can be mapped via stereographic projection onto real quaternions {\bf H},
and as in the case of complex spinors one may introduce quaternionic
spinors in terms of quaternionic homogeneous coordinates $q \to q_1/q_2$.
Note, that the quotient $q$ is mathematically well defined since real
quaternions constitute a division algebra. Depending on the metric,
sesquilinear symmetric or bilinear antisymmetric \cite{Gil}, M\"obius
transformations of the quaternionic plane give rise to transformations of
the spinor
\begin{equation}
\Psi\,=\,\left(\begin{array}{c}q_{1}\\q_{2}\end{array}\right)
\end{equation}
which can be described by the compact groups U(2,$q$) and Sp(2,$q$),
respectively. The represention of
such spinor transformations by means of complex
vector spaces is based on the isomorphisms
\begin{equation}
{\rm Sp}(2,q)\,\simeq\,{\rm USp}(4)\,\simeq\,{\rm U}(2,q)\,.
\end{equation}
A further generalization of the M\"obius transformations in the quaternionic
plane leads to Sl(2,$q$) transformations of the two quaternionic homogeneous
coordinates. The latter transformations can be realized on complex vector
spaces by the noncompact group SU$*$(4) whose algebra is isomorphic to so(5,1)
and thus generates transformations of real coordinates. In analogy to the
stereographic projection $S^{2}$ $\to$ {\bf C} where the equations of motion
are written in terms of the independent spinors $\psi \sim z$ and $\psi^{+}
\sim \overline{z}$ our approach suggests the formulation of relativistic
equations of motions in terms of the {\it independent} quaternion spinors
$\Psi \sim q$ and $\Psi^{+} \sim q^{+}$.

\noindent
These considerations throw some light on the reason for which SU(4) Wigner
supermultiplet scheme may be favoured over the relativistic SU(6) approaches
of the late 60ies solely based on the generalization of complex dimensions.
Indeed, in recalling the relations \cite{Gil}, \cite{Helga}
\begin{equation}
{\rm su}\!*(4)\,\simeq\,{\rm sl}(2,q)\,\simeq\,{\rm so}(5,1)
\end{equation}
it becomes apparent that the advantage of SU(4) over SU(6) lies
in the embedding of quaternions. Since the algebras su$*$(4) and su(4) are
related by Weyl's unitary trick, we can use the isomorphism su$*$(4)
$\simeq$ so(5,1) which leads to the deSitter group via the chain
SO(5,1) $\supset$ SO(4,1), and therefore to the Poincar\'e
group as obtained from SO(4,1) by contraction \cite{nahas}.
Thus a relativistic description of hadrons in terms of SU(4) representations
and embeddings of (twofold) quaternionic homogeneous coordinates becomes
possible.

\noindent
Note, that real quaternions have already been successfully exploited by
Finkelstein et al. \cite{finkel} for formulating quantum mechanics
consistently. We here expect quaternion `spinors' to be the most suited
mathematical tools for a quantum field theory of hadrons. For the relation
of the approach suggested to Chiral Dynamics see \cite{thesis}. In fact,
there are different approaches possible to handle the relativistic aspects
of field theories. The most elegant approach would be the construction of
quaternionic polynomials and application of an appropriate quaternionic
analysis which would also allow to study finite SO(5) rotations and the
geometry of Kaluza--Klein theories in Pauli's approach \cite{Pa 33}.
Unfortunately, due to the noncommutativity of the quaternions
and the lack of quaternionic analysis the problem has to be treated
by embedding the quaternions into complex representation spaces.

Therefore, the more realistic approach will be instead to use the SU(4)
representations and their decomposition
according to the chain SU(4) $\sim$ SO(6) $\supset$ SO(5)
in terms of real coordinates and the DKP--algebra. Alternatively, the
reduction SU(4) $\supset$ SU(2) $\times$ SU(2) \cite{dakiri}
as a complex analogon to the quaternionic
projective space {\bf HP}(1) = Sp(2)/Sp(1)$\otimes$Sp(1) \cite{Helga} can
be exploited, too.
In this respect we wish to quote early
work by Hecht and Pang \cite{hecht} where
SU(4) state vectors have been constructed
and the appropriate Wigner--Racah algebra has been worked out.
For example, the meson fields in the
vector representation [2,1,1] can be attached to
SU(4) irreducible tensors
$\label{eq:vector}
T^{[2,1,1]}_{(S m_{s})\,(T m_{t})}$,
where $(S m_{s})$ and $(T m_{t})$ denote spin and isospin quantum numbers,
respectively.
The nucleon and the delta resonance can be organized into
the [3,0,0] representation and described by means of
the totally symmetric third rank tensors
$\label{eq:spinor}
T^{[3,0,0]}_{({1\over 2} m_{s})\,({1\over 2} m_{t})}$ and
$T^{[3,0,0]}_{({3\over 2} m_{s})\,({3\over 2} m_{t})}$.
It is further possible to construct a polynomial system for SU(4)
in analogy to the spherical harmonics of the rotational group SO(3)
\cite{Mosh 63}. These polynomials are available as a basis in which
a proper SU(4) Hamiltonian can be diagonalized.
More recently, SU(4) meson--fermion vertices have been
presented in \cite{dakiri} where importance of the totally symmetric
$\lbrace 20\rbrace$--plet as intermediate
state in Born graphs for neutral pion photoproduction
on the nucleon at threshold was emphasized.
With respect to the relativistic aspects of our approach, it should be noted
that the necessary discussion of the complexified quaternionic algebra and
the associated real forms leads naturally to the last division algebra, the
octonions {\bf O}. With respect to physical ideas and recent research,
we only want to point out the deep relation between {\bf O} and the sphere
$S^{7}$. Especially the group SO(8) acting on $S^{7}$ is used in the
framework of GUT theories \cite{Collins}. Furthermore, for a thorough
discussion on the relation of octonions to quarks and the associated symmetry
groups G$_{2}$ and SU(3), we refer the interested reader to the excellent
article of G\"unaydin and G\"ursey \cite{gg}.

\section{Summary}
\setcounter{equation}{0}
In this study we propose SO(6) invariant first order wave equations
describing selfconjugate spin--flavour representations of
the group SU(4).
In the spirit of the linear $\sigma$ model we construct corresponding
effective SO(6) invariant Lagrangians on the basis of quaternionic
canonical emdeddings.
We show that the finite mass $\lbrace 4\rbrace $--plet
emerging in the SO(6) $\supset$ SO(5) $\supset$ SO(4) reduction scheme
corresponds to the hedgehog solution
of the nonlinear sigma model whereas for the
respective $\lbrace 6\rbrace$--plet
(and therefore for the relevant vector mesons degrees of freedom)
we predict a helical field configuration.
We emphasize the important r\^{o}le of quaternionic geometry and analysis
for relativistic field theory. In this context, we point out the possibility
of further generalizations of effective models not by increasing the
dimensions of their representation spaces but by additional complexification
of the underlying division algebras in the sequence {\bf C} $\to$ {\bf H}
$\to$ {\bf O} and by use of appropriate homogeneous coordinates.

\noindent
We suggest that the totally symmetric fermionic $\{20\}$--plet of
the contracted SU'(4) emerging in the large $N_c$ limit of QCD
possibly shows up as a soliton of the complexified quaternion ${\cal U}(x)$.

{\bf Acknowledgements:}
We are indebted to Prof. Karl--Friedrich Neeb for interest
and illuminating discussions on the properties of the
orthogonal groups.
This research has been partly supported by the Society
for Heavy Ion Research GSI (Darmstadt).

%\begin{references}


\begin{thebibliography}{99}
\bibitem{t'Hoo 74} G.\ 't Hooft, Nucl.\ Phys. {\bf B72}, 461 (1974)
\bibitem{Wit 79} E.\ Witten, Nucl.\ Phys. {\bf B160}, 57 (1979)
\bibitem{Be 93} R.\ K.\ Bhaduri, {\it Models of the Nucleon} (Addison--Wesley
Publ. Comp., 1988)
\bibitem{Ger 84} J.-L. Gervais and B. Sakita, Phys. Rev. D {\bf 30}, 1795
(1984)
\bibitem{Inonu} E. In\"{o}n\"{u} and E. P. Wigner, Proc. Nat. Acad. Sci.
                {\bf 39}, 510 (1953)
\bibitem{Da 93} R. Dashen and A. V. Manohar, Phys. Lett. {\bf B315}, 425
(1993); {\it ibid.} {\bf B315}, 438 (1993)
\bibitem{Ki 91} M. Kirchbach and D. O. Riska, Nuovo Cim. {\bf 104}, 1837 (1991)
\bibitem{Go 61} M. Gell--Mann and M. Levy, Nuovo Cim. {\bf 16}, 705 (1958);
J. Goldstone, {\it ibid.} {\bf 19}, 154 (1961)
\bibitem{Velo 78} G. Velo and A. S. Wightman (eds.), {\it Invariant Wave
Equations}, Lecture Notes in Physics Vol. 73 (Springer Verlag, Berlin, 1978)
\bibitem{Ro 60} H. Umezawa, {\it Quantum Field Theory} (North--Holland Publ.
Comp., Amsterdam, 1956); P.\ Roman, {\it Theory of Elementary Particles}
(North--Holland Publ. Comp., Amsterdam, 1960)
\bibitem{Pa 33} W. Pauli, Ann. d. Phys. (Leipzig) {\bf 18}, 305 (1933);
{\it ibid.} {\bf 18}, 337 (1933)
\bibitem{Bh 45} H. J. Bhabha, Rev. Mod. Phys. {\bf 17}, 200 (1945); {\it ibid.}
{\bf 21}, 451 (1949)
\bibitem{Fi 74} E. Fischbach, J. D. Louck, M. M. Nieto, and C. K. Scott,
J. Math. Phys. {\bf 15}, 60 (1974)
\bibitem{Ke 39} N. Kemmer, Proc. Roy. Soc. {\bf A173}, 91 (1939)
\bibitem{MMN} R. A. Krajcik and M. M. Nieto, Phys. Rev. D {\bf 10}, 4049
(1974); {\it ibid.} {\bf 11}, 1442 (1975); {\it ibid.} {\bf 11}, 1459 (1975);
{\it ibid.} {\bf 13}, 924 (1976); {\it ibid.} {\bf 14}, 418 (1976); {\it ibid.}
{\bf 15}, 433 (1977); {\it ibid.} {\bf 15}, 445 (1977)
\bibitem{Cornwell} J. F. Cornwell, {\it Group Theory in Physics} (Academic
Press, London, 1990)
\bibitem{Gil} R. Gilmore, {\it Lie Groups, Lie Algebras and Some of Their
Applications} (John Wiley $\&$ Sons, New York, 1974)
\bibitem{Ba 64} A. O. Barut, Phys. Rev. {\bf 135}, B839 (1964)
\bibitem{finkel} D. Finkelstein, J. M. Jauch, S. Schiminovich, and D. Speiser,
J. Math. Phys. {\bf 3}, 207 (1962)
\bibitem{Sky 62} T. H. R. Skyrme, Nucl. Phys. {\bf 31}, 556 (1962)
\bibitem{Zah 86} I. Zahed and G. E. Brown, Phys. Rep. {\bf 142}, 1 (1986)
\bibitem{Wal 93} H. Walliser, Nucl. Phys. {\bf A548}, 649 (1992)
\bibitem{Fu 87} W. I. Fushchich and A. G. Nikitin, {\it Symmetries of
Maxwell's Equations} (D. Riedel Publ. Comp., Dordrecht, 1987); W. I. Fushchich
and R. S. Shdanov, Sov. J. Part. Nucl. {\bf 19}, 1154 (1988)
\bibitem{Ok 79} S. Okubo and Y. Tosa, Phys. Rev. D {\bf 20}, 462 (1979)
\bibitem{gelf} I. M. Gel'fand, R. A. Minlos, and Z. Ya. Shapiro,
{\it Representations of the rotation and Lorentz groups and their application}
(Pergamon Press, Oxford, 1963)
\bibitem{thesis} R. Dahm, Ph. D. thesis, University Mainz, Germany, 1995
\bibitem{Helga} S. Helgason, {\it Differential Geometry and Symmetric Spaces}
(Academic Press, New York, 1962)
\bibitem{nahas} M. Levy--Nahas, J. Math. Phys. {\bf 8}, 1211 (1967)
\bibitem{hecht} K. T. Hecht and S. C. Pang, J. Math. Phys. {\bf 10}, 1571
(1969)
\bibitem{Mosh 63} M. Moshinsky, J. Math. Phys. {\bf 4}, 1128 (1963)
\bibitem{dakiri} R. Dahm, M. Kirchbach, and D. O. Riska,
                 Int. J. Mod. Phys. B9 (1995) (in press)
\bibitem{Collins} P. D. B. Collins, A. D. Martin, and E. J. Squires,
{\it Particle Physics and Cosmology} (John Wiley $\&$ Sons, New York, 1989)
\bibitem{gg} M. G\"unaydin and F. G\"ursey, J. Math. Phys. {\bf 14}, 1651
(1973)
\end{thebibliography}
\end {document}